\documentclass[runningheads]{llncs}
\usepackage[T1]{fontenc}
\usepackage{graphicx}
\usepackage{graphicx}
\usepackage{amsmath}
\usepackage{amssymb}
\usepackage{booktabs}
\usepackage{multirow}

\usepackage[breaklinks,colorlinks,linkcolor=blue,citecolor=blue,urlcolor=blue]{hyperref}
\usepackage{color}

\usepackage[capitalize]{cleveref}
\crefname{section}{Sec.}{Secs.}
\Crefname{section}{Section}{Sections}
\Crefname{table}{Table}{Tables}
\crefname{table}{Tab.}{Tabs.}
\usepackage[symbol]{footmisc}

\newcommand*\samethanks[1][\value{footnote}]{\footnotemark[#1]}

\begin{document}
%
%\title{Contribution Title\thanks{Supported by organization x.}}
\title{Deformer: Towards Displacement Field Learning for Unsupervised Medical Image Registration}

\titlerunning{Deformer}
%\author{First Author\inst{1}\orcidID{0000-1111-2222-3333} \and
%Second Author\inst{2,3}\orcidID{1111-2222-3333-4444} \and
%Third Author\inst{3}\orcidID{2222--3333-4444-5555}}
%

\author{Jiashun Chen\inst{1}\thanks{Equal contribution and the work was done at Tencent Jarvis Lab.} \and
Donghuan Lu\inst{2}\samethanks[1] \and
Yu Zhang\inst{1}\thanks{Y. Zhang and D. Wei are the corresponding authors.} \and
Dong Wei\inst{2}\samethanks[2] \and
Munan Ning\inst{2} \and
Xinyu Shi\inst{1} \and
Zhe Xu\inst{3} \and
Yefeng Zheng\inst{2}
}
%1{Jiashun Chen}
%2{Donghuan Lu}
%3{Yu Zhang}
%4{Dong Wei}
%5{Munan Ning}
%6{Xinyu Shi}
%7{Zhe Xu}
%8{Yefeng Zheng}

\authorrunning{J. Chen et al.}
\institute{School of Computer Science and Engineering, Southeast University, Nanjing, China \\
\email{zhang\_yu@seu.edu.cn} \and
Tencent Healthcare Co., Jarvis Lab, Shenzhen, China \\
\email{donwei@tencent.com} \and
Biomedical Engineering, The Chinese University of Hong Kong, Hong Kong, China \\
}

\maketitle              % typeset the header of the contribution
\begin{abstract}
%The abstract should briefly summarize the contents of the paper in 150--250 words.
Recently, deep-learning-based approaches have been widely studied for deformable image registration task. However, most efforts directly map the composite image representation to spatial transformation through the convolutional neural network, ignoring its limited ability to capture spatial correspondence. On the other hand, Transformer can better characterize the spatial relationship with attention mechanism, its long-range dependency may be harmful to the registration task, where voxels with too large distances are unlikely to be corresponding pairs. In this study, we propose a novel Deformer module along with a multi-scale framework for the deformable image registration task. The Deformer module is designed to facilitate the mapping from image representation to spatial transformation by formulating the displacement vector prediction as the weighted summation of several bases. With the multi-scale framework to predict the displacement fields in a coarse-to-fine manner, superior performance can be achieved compared with traditional and learning-based approaches. Comprehensive experiments on two public datasets are conducted to demonstrate the effectiveness of the proposed Deformer module as well as the multi-scale framework.

\end{abstract}

%\section{First Section}
%\subsection{A Subsection Sample}
%Please note that the first paragraph of a section or subsection is
%not indented. The first paragraph that follows a table, figure,
%equation etc. does not need an indent, either.

%Subsequent paragraphs, however, are indented.

%\subsubsection{Sample Heading (Third Level)} Only two levels of
%headings should be numbered. Lower level headings remain unnumbered;
%they are formatted as run-in headings.

%\paragraph{Sample Heading (Fourth Level)}
%The contribution should contain no more than four levels of
%headings. Table~\ref{tab1} gives a summary of all heading levels.

% the environments 'definition', 'lemma', 'proposition', 'corollary',
% 'remark', and 'example' are defined in the LLNCS documentclass as well.

\section{Introduction}
\label{sec:intro}
%Start with formulate the displacement vector as the weighted average of the displacement basis.Unlike previous studies using U-Net architecture, which extract feature maps at different scales, we direct extract displacement vector fields at different scales, so that auxilary losses can be applied on these fields as well. Previous methods only leverage the learning ability of deep learning methods, without sufficient guidance from prior knowledge and large enough amount of data, the performance is limited. attention is useful but limited. treat as segmentation task, ignoring the intrinsic property of registration task.
Deformable image registration (DIR), which aims to estimate a proper deformable field $\phi$ that can warp the moving image $I_{m}$ to align with the fixed image $I_{f}$, is an essential procedure in various medial image analysis tasks, such as surgical navigation~\cite{gou2018large}, image reconstruction~\cite{li2010real} and atlas construction~\cite{dalca2019learning}. Traditional registration approaches~\cite{avants2008symmetric,vercauteren2009diffeomorphic,wang2005validation} align voxels with similar appearance through solving an optimization problem for each volume pair. Unfortunately, the computational intensive optimization limits their usage in practical clinical applications.

Recently, unsupervised deep-learning-based DIR approaches~\cite{balakrishnan2019voxelmorph,xu2020adversarial,hu2019dual,xu2021double,xu2021f3rnet} have been widely studied for their computational efficiency. Many methods, such as VoxelMorph~\cite{balakrishnan2019voxelmorph}, Dual-PRNet~\cite{hu2019dual} and FAIM~\cite{kuang2019faim}, adopt the convolutional neural network (CNN) as their backbone because of its superior performance for various vision tasks~\cite{huang2017densely,tan2019efficientnet}. However, CNN shows limited capability in capturing spatial relationship~\cite{song2021cross}, which becomes a bottleneck for these methods. With the recent development of vision Transformer, some efforts have also been made to explore its effectiveness in DIR task~\cite{chen2021vit}. Although its attention mechanism~\cite{song2021cross} is potentially more suitable to characterize spatial relationship, directly applying it for DIR task may lead to inferior performance due to its long-range dependency. Because of the fixed anatomical structure of medical scans, preserving tissue discontinuity~\cite{chen2021deep} is essential for medical image registration task. Therefore, the corresponding voxel should only be found in a limited local range, which could be undermined by the long-range dependency. 

%\begin{figure}[t] 
%\centering
%\includegraphics[width=0.6\columnwidth]{Fig/intro.pdf} 
%\caption{The proposed Deformer module utilizes two linear layers to learn the displacement basis and the attention weight to compose the displacement vector. 'C' represents concatenation of feature maps, and '*' denotes matrix multiplication. }
%\caption{Compared with traditional transformer, the Deformer utilizes two linear layers to mimic the displacement basis and the attention weight to obtain the displacement field, which focuses on local information among the displacement basis instead of long range dependence.}
%\label{fig:intro} 
%\end{figure}

To this end, we propose a Deformer module to explicitly exploit the intrinsic property of registration task for facilitating the mapping from image representation to spatial transformation. We make a simple yet critical formulation that the displacement between a pair of voxels can be considered as the weighted summation of several basic vectors (referred as displacement bases as well) with three elements, representing the $x$, $y$ and $z$ components, respectively. Leveraging attention mechanism's ability to capture spatial relationship from image representation, the proposed Deformer module adopts two separate branches to implement such a paradigm. The first branch learns the displacement bases, denoting the potential deformable directions, while the other one predicts the attention weight for each basis, representing the offset length along each direction. Thus, the voxel-wise displacement can be obtained via matrix product of the basis vectors and the attention weights. In addition, the multi-head strategy is applied to enable the Deformer module to extract latent information from different representation subspaces. Furthermore, a multi-scale framework, namely Deformer-based Multi-scale Registration (DMR) is customized to further boost the performance. Unlike previous methods, which either successively predict the displacement fields at different scales~\cite{avants2008symmetric,mok2020large,wang2005validation} or only exploit multi-scale feature maps through U-Net architecture~\cite{balakrishnan2019voxelmorph,rohe2017svf}, we propose to introduce the Deformer module along with an auxiliary loss at each scale to learn the displacement fields in a coarse-to-fine manner and automatically fuse them through the tailored refining network, such that the information at different scales can be fully exploited without intensive successive computation.

\begin{figure}[hb] 
\centering 
\includegraphics[width=\columnwidth]{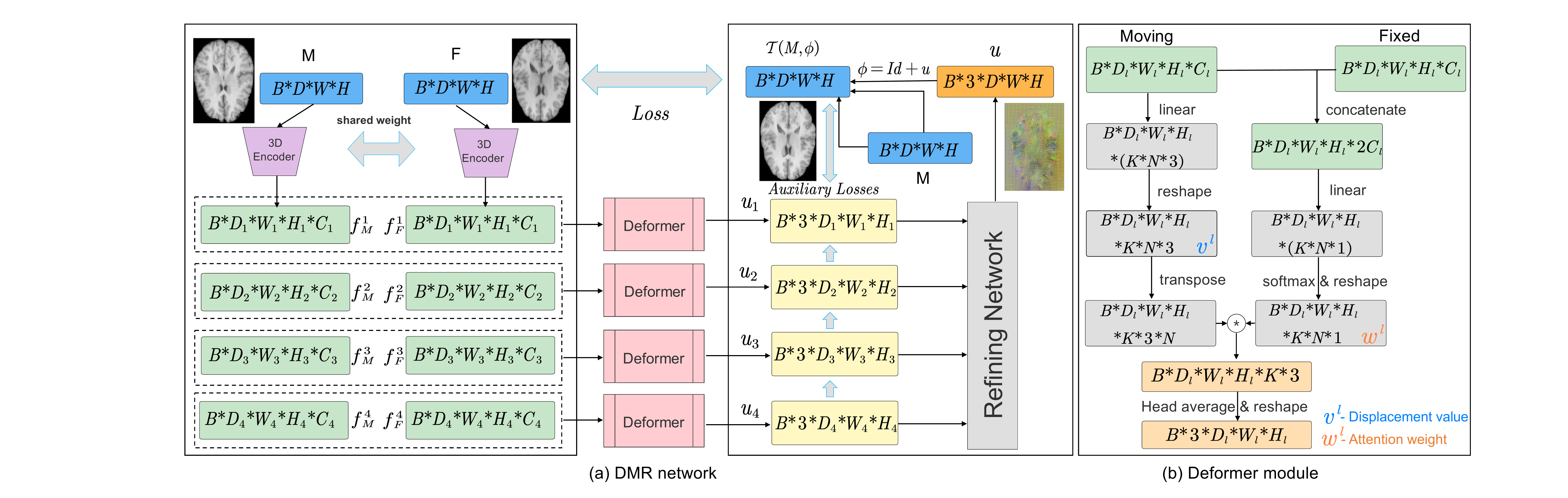} 
%\caption{The overview of the proposed DMR framework and the Deformer module. (a) The DMR network consists of multi-scale encoder to extract $L$ pairs of feature maps from input moving and fixed 3D volumetric scans, the Deformer module to obtain a displacement field at each scale and a refining network to fuse multi-scale displacement fields into a final displacement field. Additional penalties, denoted as auxiliary losses, are imposed on the intermediate displacement fields at each scale. (b) The architecture of the Deformer module is composed of two branches with fully connected layers. The displacement basis is learned by a linear layer of the left branch with the feature map of moving scan, while the attention weight for each displacement basis is monitored by a linear layer of the right branch with the concatenated feature map from both the moving and fixed images.}
\caption{The proposed DMR framework (a) and Deformer module (b).} %(a) The DMR network consisting of multi-scale encoder, the Deformer module and a refining network. Additional penalties, denoted as auxiliary losses, are imposed on the intermediate displacement fields at each scale. (b) The architecture of the Deformer module is composed of two branches with fully connected layers. The displacement basis is learned by a linear layer of the left branch with the feature map of moving scan, while the attention weight for each displacement basis is monitored by a linear layer of the right branch with the concatenated feature map from both the moving and fixed images.}
\label{Fig.overview} 
\end{figure}
%------------------------------------------------------------------------

\section{Method}
\label{sec:method}

\subsubsection{Problem Setting.}
\label{sec:problem}
Given a pair of moving and fixed medical 3D scans (called images or volumes as well) $\{M,F\}\in\mathbb{R}^{D\times W \times H}$, the objective of image registration is to estimate a deformation field $\phi\in\mathbb{R}^{3\times D\times W\times H}$ such that the warped moving scan $\mathcal{T}(M,\phi)\in\mathbb{R}^{D\times W\times H}$ can be aligned with the fixed scan $F$, where $D$ denotes the depth, $W$ the width, $H$ the height and $3$ for the 3D spatial dimension. %The registration process can be considered as finding the most similar voxel in the fixed image for each voxel in the moving image and the mapping function between two images is often represented by the deformation field $\phi$ of displacement vectors between each pair of voxels. 
Specifically, the $\phi$ can be represented as $Id+u$, where $Id$ denotes identity transformation and $u$ represents the displacement field. To be consistent with previous registration studies~\cite{balakrishnan2019voxelmorph,chen2021vit,kim2021cyclemorph}, a spatial transformation network~\cite{jaderberg2015spatial} is adopted to map the moving image to the fixed image with the estimated displacement field. 
%To clarify notations, $L$ represents the total number of scales, while $\{f_M^l,f_F^l\}$ denotes a pair of moving and fixed feature maps extracted by the encoder at scale $l$. $K$ is the number of heads in each Deformer module and $N$ represents the number of displacement basis in each head. $B$ denotes the batch size, while $C_l$ and $u^l$ represent the channel of feature map and the intermediate displacement field obtained by Deformer module at scale $l$, respectively. %output channel of a 3D convloution block  which raises the channel of sub-flow fields before fed to the decoder, $u^l$ denote the sub-flow field obtained by Deformer at scale $l$, $\{u_l^{l-1}\}$ express the fused flow field combining scale $l$ and $l-1$ in \cref{sec:decoder}; $\Omega$ is the total voxel set, for each voxel $p$, $p_i$ include its $n^3$ neighbour voxel; both $\overline{F(p_i)}$ and $ \overline{\mathcal{T}(M,\phi)(p_i)}$ represents the average function value of $n^3$ neighbour voxel; $\bigtriangledown$ is an approximate spatial gradients using differences between neighboring in \cref{sec:loss function}.

\subsubsection{Method Overview.}
As shown in Fig.~\ref{Fig.overview}, the proposed DMR framework consists of three main components, the multi-scale encoder to extract L ($L=4$ in this study) pairs of feature maps $\{f_M^l,f_F^l\}_{l=1}^L$ from both the moving image $M$ and the fixed image $F$, the proposed Deformer modules to exploit the interaction between $M$ and $F$ to deliver the displacement vector fields from coarse to fine, and the refining network to combine the latent information from different representation subspaces to enable high-resolution large-deformable registration. The detailed explanations of the Deformer module, the DMR network, as well as the loss function are stated in the following sections. % Note that other than the objective function imposed on the final flow field, we propose to impose auxiliary loss functions on each intermediate flow fields with different scales. 

%\subsection{Multi-scale Encoder}
%\label{sec:encoder}

\subsubsection{Deformer Module.}
To facilitate the mapping of the image representation to spatial relationship and increase the interpretability of the network, a Deformer module is designed to learn the displacement field. Based on the linear algebra theory, a displacement vector between two voxels can be formulated as the weighted summation of several basic vectors. For implementing this paradigm through network, the proposed Deformer module adopts attention mechanism with two separate branches, as displayed in Fig.~\ref{Fig.overview}b. %to obtain the displacement filed $u_l$ from the extracted feature maps at each scale $l$. 

The first branch, i.e., the left branch in Fig.~\ref{Fig.overview}b, adopts a linear projection to convert every voxel-wise feature vector into $N$ bases, each of which consists of three elements to represent the displacement values in three spatial directions, i.e., the $x$, $y$ and $z$ axes, respectively. In addition, we adopt the multi-head strategy~\cite{vaswani2017attention} so that the information from different representation subspaces can be obtained, leading to $K\times N$ bases for each voxel, where $K$ represents the number of heads in each Deformer module. Note that only the feature maps of the moving images $f_M^l$ are employed in this branch, which is sufficient to extract the most representative displacement bases for denoting the most likely deformation directions as demonstrated by the ablation study. On the other hand, the second branch, i.e., the right branch in Fig.~\ref{Fig.overview}b, adopts the concatenated feature map $[f_M^l, f_F^l]$ from both the moving and fixed images to learn the attention weight for each displacement basis. The similarity between the moving image and the fixed image is measured by a linear layer followed by a softmax function to impose the non-linearity of the module. It is worth mentioning that both linear projections are performed on each pair of voxel-wise feature vectors independently, resulting in a small number of parameters comparing with standard CNN or Transformer block. Besides, the latent information of the nearby voxels has already been incorporated into the feature vector through the CNN-based encoder. Therefore, the similarity measurement is not limited to the voxels at the same location, and the field-of-view is expanded with the decreasing of the feature map resolution. Finally, the attention weights are multiplied to the displacement bases followed by a head-wise average to obtain the displacement field $u^l$ at scale $l$. The overall process can be formulated as:
\begin{equation}
u^l=\frac{1}{K}\sum_{i=1}^K\sum_{j=1}^N w_{i,j}^l\times v^l_{i,j}, \quad l=1,...,L,
\label{eq:displacement field}
\end{equation}
\begin{equation}
v^l_{i,j}=fc_1(f_M^l), \quad w_{i,j}=fc_2([f_M^l, f_F^l]),
\label{eq:head}
\end{equation}
where $v^l_{i,j}$ denotes the displacement basis, $w_{i,j}$ is the attention weight, $[\cdot]$ represents the concatenation operation, while $fc_1$ and $fc_2$ denote the mapping functions of two branches, respectively.

%Given the multi-scale feature maps extracted by different CNN blocks, we adopt $L$ Deformer modules corresponding to each of them to obtain the displacement fields in different scales. With the increasing of scales, $L$ displacement fields are obtained from coarse to fine.

\subsubsection{Network Architecture.}
\label{sec:refine}
In order to fully exploit the latent information at different representation subspaces without introducing the intensive computation of successive network, we propose a DMR framework to learn displacement fields at different scales via the proposed Deformer module along with auxiliary loss to provide additional guidance. 

As shown in Fig.~\ref{Fig.overview}, we first adopt a multi-scale CNN encoder to extract the latent representations at different scales from both the moving and fixed images. Specifically, sharing a similar architecture as ~\cite{balakrishnan2019voxelmorph}, the encoder is composed of $L$ 3D convolution blocks, each of which consists of a convolutional layer with a stride of 2 and a kernel size of 4, and a batch normalization layer followed by a leaky rectified linear unit (LeakyReLU) with a negative slope of 0.2. Therefore, the encoder can extract a sequence of $L$ pairs of intermediate feature maps $\{f_M^l,f_F^l\}_{l=1}^L$ with different scales through different convolution blocks. Note that both the moving and the fixed images are fed into the same encoder for feature extraction.

%Unlike most previous studies, which directly feed the extracted features to the decoder through short connections, we propose to introduce Deformer modules to learn the displacement fields from the pairs of feature maps at each scale. Auxiliary loss is imposed on the output of each Deformer module to provide additional guidance for fully exploiting the information from representation subspace at each scale. Finally, a refining network is applied to exploit the information of displacement fields and the pair of image feature maps at multiple scales of abstraction to deliver high-resolution large-deformation mappings.

The refining network consists of repeated application of the fusion block followed by a convolution head to convert the last feature map to the final displacement field at the original resolution.  Specifically, we denote the output feature map of fusion block $l$ as $g^l$, then the displacement field $u^l$ at scale $l$ is converted to a feature map $h^l$ with a fixed number of channels $C$ via a convolution block. Subsequently, the feature map of the last fusion block $g^{l+1}$ is upsampled by a bilinear interpolation and added to $h^l$ followed by the concatenation of the latent representations $f_M^l$ and $f_F^l$ to provide image information, formulated as:

%Each block contains three convolutional layers with $C_{r1}$, $C_{r2}$ and $C$ channels, respectively, each of which has a stride of 1 followed by a group normalization and a ReLU activation function. Note that the convolution blocks at different scales do not share weights and their kernel sizes are set as (5,5,5), (5,5,3), (5,3,3), (3,3,3) from large to small scale. 
\begin{equation}
  g^l=
  \begin{cases}
  h^l, \quad l=L
  \\\left [ 2\times upsample(g^{l+1})+h^{l},f_M^{l},f_F^{l} \right ],otherwise.
  \end{cases}
\label{eq:fused u}
\end{equation}

The final component of the fusion block is another convolution block, i.e., Conv3d reducer, for reducing the channel of output high-level latent representations to $C$. After three cascaded fusion blocks, a convolution head including three decoding blocks, each of which has two successive structures with a convolutional layer and a ReLU activation function, is applied to deliver the final displacement field. An upsample layer is introduced after the first decoding block to restore the original resolution. More details about the refining network can be found in the supplementary materials.

\subsubsection{Loss function.}
\label{sec:loss function}
To validate the effectiveness of the proposed DMR framework and Deformer module, we adopt the most commonly used objective function~\cite{balakrishnan2019voxelmorph} for a fair comparison, which is defined as:
\begin{equation}
  \mathcal{L}(M,F,\phi)=\mathcal{L}_{sim}(\mathcal{T}(M,\phi), F)+\lambda\mathcal{L}_{reg}(\phi),
  \label{eq:loss0}
\end{equation}
where $\lambda$ is the regularization trade-off parameter, setting as 1 empirically. The first term $\mathcal{L}_{sim}(\mathcal{T}(M,\phi), F)$ measures the similarity between the warped moving images and the fixed scans using local normalized cross-correlation. The second term $ \mathcal{L}_{reg}(\phi)$ is a regularization imposed on the displacement fields to penalize local spatial variation. %Both terms are consistent with Voxelmorph~\cite{balakrishnan2019voxelmorph}.

%The first term $\mathcal{L}_{sim}(\mathcal{T}(M,\phi), F)$ measures the similarity between the warped moving images and the fixed scans using local normalized cross-correlation, which can formulated as:
%\begin{equation}
%\mathcal{L}_{sim}(\mathcal{T}(M,\phi), F) =\sum_{p\in\Omega}\frac{(\sum\limits_{q\in P}\Delta F(q) \Delta %\mathcal{T}(q))^2}{\sum\limits_{q\in P}(\Delta F(q))^2\sum\limits_{q\in P}(\Delta \mathcal{T}(q))^2},
%  \label{eq:sim}
%\end{equation}
%where $\Omega$ is the entire voxel set, and $P$ denotes the $n^3$ neighbours around voxel $p$. If we use $\mathbb{E}(F(p))$ and $\mathbb{E}(\mathcal{T}(M,\phi)(p))$ to represent the value expectation of $n^3$ neighboring voxels on the fixed images and the warped images, $\Delta F(q)$ and $\Delta \mathcal{T}(q)$ can be defined as $F(q)-\mathbb{E}(F(p))$ and $\mathcal{T}(M,\phi)(p)-\mathbb{E}(\mathcal{T}(M,\phi)(p))$, respectively.

%The second term is a regularization imposed on the displacement fields to penalize local spatial variation, calculated by:

%\begin{equation}
%\mathcal{L}_{reg}(\phi)=\sum_{p\in \Omega }\left \| \nabla u(p)  \right \| ^2,
%  \label{eq:reg}
%\end{equation}
%where $\nabla=(\nabla_x, \nabla_y, \nabla_z)$ is an approximate spatial gradients of displacement field values between two neighboring voxels. $\nabla_x u(p)$ is computed by $u(p_{(x+1)},p_y,p_z)-u(p_x,p_y,p_z)$ and similar operation is applied to obtain  $\nabla_y u(p)$ and  $\nabla_z u(p)$.

As stated above, we propose to impose penalty on the intermediate displacement field at each scale to provide direct guidance for each level of Deformer module as well as the feature extractor. It is worth mentioning that the intermediate displacement fields should be upsampled to the original scale to warp the moving image for the computation of objective function. The overall loss function can be written as:
\begin{equation}
  \mathcal{L}_{total}=\mathcal{L}(M,F,\phi)+\sum_{l=1}^L\beta_{l}\mathcal{L}(M,F,\phi^l),
  \label{eq:loss}
\end{equation}
where $\beta_{l}$ is the weight for the intermediate loss function at scale $l$.

%------------------------------------------------------------------------------------
\section{Experiments and Discussion}
\label{sec:experiment}
\subsubsection{Datasets.}
\label{sec:dataset and metric}
To validate the performance of the proposed approach, we conduct experiments on two publicly available datasets, i.e., the LONI Probabilistic Brain Atlas (LPBA40) dataset~\cite{shattuck2008construction} and the Neurite subset of the Open Access Series of Imaging Studies (Neurite-OASIS) dataset~\cite{marcus2007open}. %{\color{red} why only subset. The dataset is from learn2reg 2021 challenge. The deformation field is more local and the displacement value is smaller than the original OASIS dataset. Therefore, this subset is more difficult for registration and is a challenge. }
The LPBA40 dataset~\cite{shattuck2008construction} contains 40 3D volumes of whole-brain Magnetic Resonance Imaging (MRI) scans from normal volunteers, which is divided into 29, 3 and 8 scans for training, validation and testing, respectively. Skull stripping and spatial normalization are performed on each scan, followed by padding and cropping to ensure the same size of 160$\times$192$\times$160 voxels for each scan. Manual annotation of 56 structures are provided as the ground truth. The Neurite-OASIS dataset\footnote{https://learn2reg.grand-challenge.org/Datasets/} is from the learn2reg 2021 challenge \cite{hering2021learn2reg} and is a part of the OASIS Dataset Project~\cite{marcus2007open}, It contains 414 inter-patient 3D T1-weighted MRI brain scans from abnormal subjects with various stages of cognitive decline, which is split into 374, 20 and 20 scans for training, validation and testing, respectively. The scans are pre-processed by the challenge organizer, including skull stripping and spatial normalization via FreeSurfer and SAMSEG, and then cropped to 224$\times$192$\times$160 voxels. The subcortical segmentation maps of the 35 anatomical structures are provided as the ground truth for evaluation.

\subsubsection{Evaluation Criteria.}
In the experiments of both datasets, we use subject-to-subject registration for optimization, where each pair of volume is selected randomly from the training sets. For evaluation, 8 LPBA40 and 20 Neurite-OASIS scans are mapped to a standard atlas~\cite{avants2008symmetric}. %Because there is no direct measurement for the performance of registration, 
Following previous works~\cite{balakrishnan2019voxelmorph,kim2021cyclemorph}, we adopt the commonly used average Dice score of the region-of-interest (ROI) masks between the warped images and fixed images as the main evaluation metric. To quantify the diffeomorphism and smoothness of the deformation fields, the average percentage of voxels with non-positive Jacobian determinant ($|J_{\phi}| \leq 0$) in the deformation fields, the standard deviation of the Jacobian determinant ($std(|J_{\phi}|)$) the number of parameters, GPU memory and average running time to register each pair of scans on Neurite-OASIS dataset are also provided as supplementary metrics.% It is worth mentioning that Dice is the main evaluation criterion, while the other two metrics are supplementary for measuring folding during image warping, and a meaningless identity mapping achieves the lowest scores in the latter two metrics.

\begin{figure}[t]
  \centering
   \includegraphics[width=1.0\columnwidth]{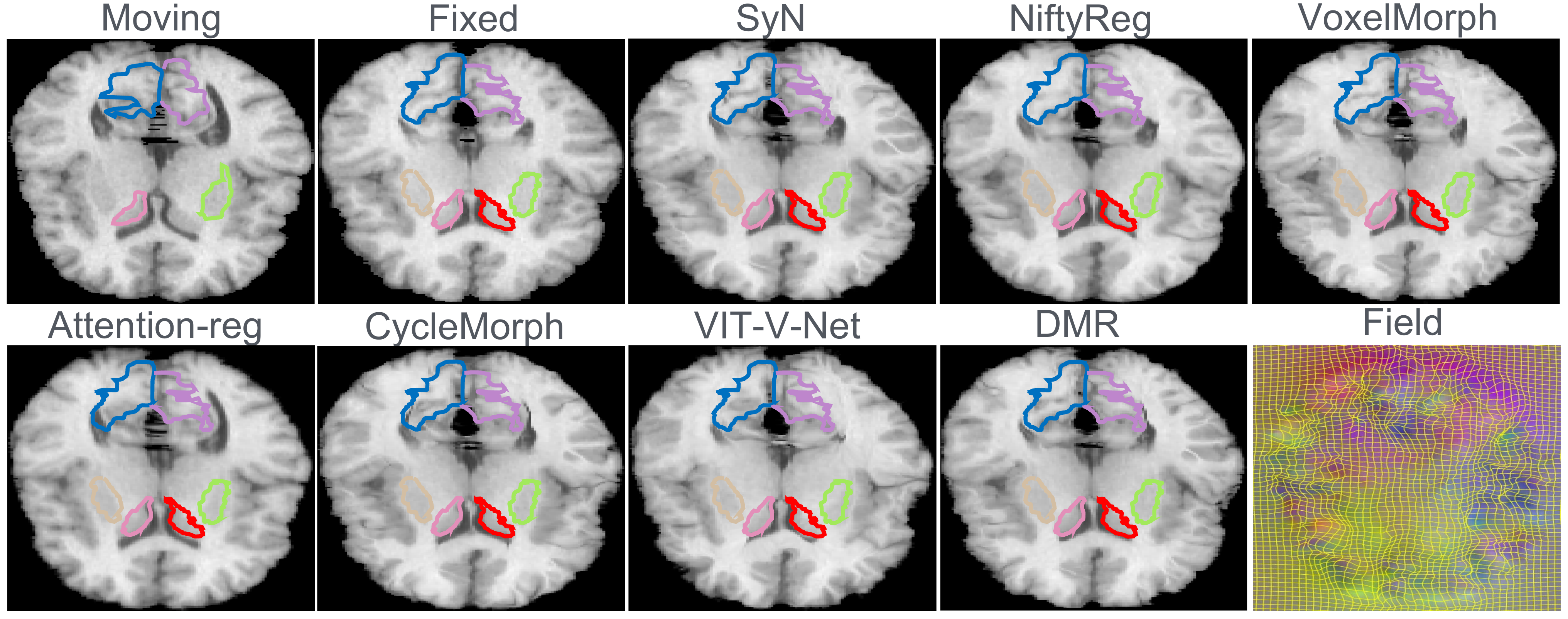}
   \caption{Qualitative results with segmentation labels of the example axial MR slices from the moving, fixed and warped images from different methods. The color curves represent the boundaries of several structures, including caudate (pink/red), putamen (brown/green), and lingual gyrus (blue/purple). The last image in the second row is the visualization of the final displacement field of our method.}
   \label{fig:qualitative results}
\end{figure}

\subsubsection{Implementation Details.}
\label{sec:implementation details}
Our method is implemented with PyTorch 1.4 and optimized using Adam optimizer~\cite{kingma2014adam} with mini-batch stochastic gradient descent. The model is trained on 4 NVIDIA V100 GPUs for 2,000 epochs. The batch size is set as 4 pairs. The learning rate is set as $4\times10^{-4}$ for the LPBA40 dataset, and $10^{-3}$ for the Neurite-OASIS dataset to reduce the time consumption as it contains more images with larger size compared to LPBA40. Similar to previous studies~\cite{balakrishnan2019voxelmorph,mok2020large}, no data augmentation is employed in our experiments. As for the Deformer modules, the head number $K$ is $8$ for all scales and the number $N$ of displacement bases for each head is 64. In the refining network, $C$ is set to 128. %the convolution block includes three convolutional layers with $C_{r1}=16$, $C_{r2}=64$, $C=128$ channels, respectively. 
For optimization, we measure image similarity using local normalized cross-correlation with windows size of $9 \times 9 \times 9$ in all the loss functions. The regularization parameter $\lambda$ is empirically set as 1.0 and the weights $\beta_l$ are all set to 1. Our implementation is publicly available at \href{https://github.com/CJSOrange/DMR-Deformer}{https://github.com/CJSOrange/DMR-Deformer}.

\subsubsection{Comparison Study.}
\label{sec:baseline methods and results}
For a quantitative comparison, we first compare with five state-of-the-art methods. Two of them are traditional methods, i.e., SyN~\cite{avants2008symmetric} and NiftyReg~\cite{modat2010fast}, two of them utilize convolution neural networks, including VoxelMorph~\cite{balakrishnan2019voxelmorph} and CycleMorph~\cite{kim2021cyclemorph}, and the rest two approaches incorporate vision Transformer or attention mechanism into the network, i.e., VIT-V-Net~\cite{chen2021vit} and Attention-reg~\cite{song2021cross}. The results of these approaches are obtained based on the official codes released by the authors. As shown in Table~\ref{tab:result1}, the proposed approach achieves 68.4\% in Dice on the LPBA40 dataset and 80.4\% in Dice on Neurite-OASIS , which outperforms the second best methods by 1.5\% and 1.6\%, respectively. Our method only generates a small percentage of folding voxels (0.624\% and 1.024\%) on both datasets, indicating reasonable smooth deformation fields. Moreover, we can reduce last three metrics to 7.2M/46.7G/0.61s by using a single head with little performance degradation, as shown in Table 1 of the supplementary. Fig.~\ref{fig:qualitative results} illustrates the registration results of various methods for qualitative analysis. Similar to the quantitative comparison, the proposed DMR achieves the most appealing qualitative results with better alignment between the warped image and the anatomical structure boundaries. For comparing with other multiple cascaded networks on Neurite-OASIS, we add a weakly-supervised Dice loss. DMR achieves 84.2\% Dice which is comparable to Learn2Reg winner LapIRN~\cite{mok2020large} (86.2\% Dice), and is superior to DLIR~\cite{de2019deep} (82.9\%) and mIVIRNET~\cite{hering2019mlvirnet} (83.4\%). For detailed evaluation of the Dice scores for individual anatomical structures and more visualization of segmentation results as well as the displacement fields, please refer to the supplementary materials.    %The reason for the relative small improvement on Neurite-OASIS may be because the samples in this dataset are all abnormal.  Compared with LPBA40, due to the presence of lesion area, the difference between the paired images is relatively large, which brings great challenges to registration.} 
%In addition, we present the visualization of the displacement fields at different scales in Fig.~\ref{fig:field}. It supports our assumption that the displacement fields at coarse scales can capture large deformation, while the ones at fine scales can provide precise transformation at high resolution, and the refining network can combine their information to enable high-resolution large-deformable registration.
%For all methods, we use their official implementation online with default parameters and change the training mode to "scan to scan". Specially,  we change the decoder to a deformable version for attention-reg. All methods are trained from scratch. Since only LapIRN used the multi-resolution cascade strategy, we compared it separately, including the results of full-resolution and three-resolution cascades.

\begin{table*}[h]
  \centering
  \caption{Comparison results of the proposed DMR framework and other state-of-the-art methods on LPBA40~\cite{shattuck2008construction} and Neurite-OASIS~\cite{marcus2007open}.}
    \scalebox{0.77}{
    \begin{tabular}{cccccccccc}
    \toprule
    \multirow{2}[3]{*}{Method} & \multicolumn{3}{c}{LPBA40} & \multicolumn{6}{c}{Neurite-OASIS} \\
\cmidrule{2-10}          & Dice (\%) & \multicolumn{1}{c}{$|J_{\phi}|\leq0$ (\%)} & \multicolumn{1}{c}{std($|J_{\phi}|$)} & Dice (\%) & \multicolumn{1}{c}{$|J_{\phi}|\leq0$ (\%)} & \multicolumn{1}{c}{std($|J_{\phi}|$)} &
\multicolumn{1}{c}{Para(M)} &
\multicolumn{1}{c}{Memo(G)} &
\multicolumn{1}{c}{Time(s)}
\\
\midrule
SyN~\cite{avants2008symmetric}   &   66.5    &   0    &   0.126    &    78.0   &   0    & 0.124  & - & - & 1504\\
NiftyReg~\cite{modat2010fast} &  66.9    & 0.135      &    0.093   &  78.5     &    0.102   &  0.197 & - & - & 378\\
VoxelMorph~\cite{balakrishnan2019voxelmorph} & 64.2  &   0.961    &  0.379   & 78.1  &   1.236    & 0.463 & 0.573 & 34.9 & 0.56\\
CycleMorph~\cite{kim2021cyclemorph} & 65.0  &  0.437    &  0.216  &  78.8  &  0.854 & 0.377 &0.784 & 42.8 & 0.58 \\
VIT-V-Net~\cite{chen2021vit} &  61.3 &  1.307     & 0.481     &  78.2 &     2.045  &  0.902 & 31.6 & 60.5 & 0.85\\
Attention-reg~\cite{song2021cross} & 62.7  & 0.808      &  0.342       &  77.5 &  1.435   & 0.501 & 0.883 & 56.4 & 0.78\\
\midrule
DMR (Ours) & \textbf{68.4}  &  0.624     &    0.334   & \textbf{80.4}  &  1.024     & 0.441 &7.9 & 120.3 & 0.63\\
\bottomrule
    \end{tabular}}%
  \label{tab:result1}%
\end{table*}%

\subsubsection{Ablation Study.}
\label{sec:ablation study}
For ablation study, we first conduct experiments to verify the effectiveness of the proposed Deformer module along with the auxiliary losses. Specifically, we compare the Deformer module with four variants: 1) Deformer-A: feeding the concatenated feature maps to the left branch for displacement basis extraction, instead of using only the feature maps of the moving images; 2) Deformer-B: removing the left branch of the Deformer module, which is equivalent to fix the displacement bases as (0, 0, 1), (0, 1, 0) and (1, 0, 0); 3) CNN: replacing the Deformer module with a CNN block, i.e., 3D ResNet~\cite{hara2018can}; 4) Transfomer: replacing the Deformer module with a vision Transformer block~\cite{dosovitskiy2020image}. It is worth mentioning that with only two fully connected layers performed on voxel-wise feature vectors, the proposed Deformer module has fewer parameters compared with CNN or Transformer (7.98M vs. 12.35M and 132M). The results in Table~\ref{tab:basis} demonstrate that the proposed Deformer module can better characterize the spatial transformation from image representation and using the moving images alone is sufficient to find optimal displacement bases.
We further evaluate the impact of Deformer modules at different scales by gradually removing the Deformer modules from fine to coarse scales ($1/2$, $1/4$, $1/8$ and $1/16$). Naturally, the auxiliary loss at the same scale is removed along with the Deformer module. As shown in Table~\ref{tab:scale number}, with the removing of Deformer modules, we can observe steadily degeneration of registration performance from 68.4\% to 66.2\% in Dice on the LPBA40 dataset, supporting the assumption that the multi-scale Deformer module and the corresponding auxiliary loss can effectively improve the registration ability of the network. The impact of more hyper-parameters can be found in the supplementary materials. %Thus, it is obvious that Deformer play an important role for extract semantic displacement information from the paired feature maps. Besides, we assumed that shallow features contain more base information about the original pixel and can be flexibly learned by our proposed deformer at the original resolution level. Besides, we also replace all the Deformer module with VIT or convolution network, e.g., 3D ResNet. Limited to memory, we only take the last three scales to replace. Rather than the fully connection layer, a convolution head with channel of 3 is connected to VIT~\cite{dosovitskiy2020image} and 3D ResNet~\cite{hara2018can} to output the displacement field of this scale. The results in \cref{tab:scale number} shows that our Deformer module performs better than VIT and 3D ResNet.

\begin{minipage}{\textwidth}
\begin{minipage}[t]{0.45\textwidth}
\makeatletter\def\@captype{table}
\setlength{\belowcaptionskip}{10pt}
\centering
  \caption{Evaluation of different conversion modules on LPBA40~\cite{shattuck2008construction}, including Deformer-A, Deformer-B, CNN and Transformer.}
  \scalebox{0.7}{
    \begin{tabular}{cccc}
    \toprule
%\multirow{2}[3]{*}{Method} & \multicolumn{3}{c}{LPBA40} \\
%\cmidrule{2-4}  & Dice (\%) & $|J_{\phi}|\leq0 (\%)$ & \multicolumn{1}{c}{std($|J_{\phi}|$)} \\
Method & Dice (\%) & $|J_{\phi}|\leq0 (\%)$ & std($|J_{\phi}|$) \\
    \midrule
    Deformer (Ours)     & 68.4  & 0.624 & 0.334\\
    Deformer-A (Ours)   & 68.0  & 0.702 & 0.368 \\
    Deformer-B (Ours)   & 66.2  & 0.945 & 0.432\\
    CNN~\cite{hara2018can}   & 66.7   & 0.896    & 0.426  \\
    Transformer~\cite{dosovitskiy2020image} & 66.3 & 1.003    & 0.455\\
    \bottomrule
    \end{tabular}}%

  \label{tab:basis}%
\end{minipage} \qquad
\begin{minipage}[t]{0.4\textwidth}
\makeatletter\def\@captype{table}
\setlength{\belowcaptionskip}{10pt}
\centering
  \caption{The impact of the number of Deformer modules on LPBA40~\cite{shattuck2008construction}. ``$\checkmark$" represents applying the Deformer module at this scale.}
  \scalebox{0.65}{
    \begin{tabular}{rrrrcrr}
    \hline
    \multicolumn{4}{c}{Scale}     & \multicolumn{3}{c}{LPBA40} \\
\cmidrule{1-7}    \multicolumn{1}{c|}{1/2}  & \multicolumn{1}{c|}{1/4} & \multicolumn{1}{c|}{1/8} & \multicolumn{1}{c|}{1/16} & Dice (\%) & \multicolumn{1}{c}{$|J_{\phi}|\leq0 (\%)$} & \multicolumn{1}{c}{std($|J_{\phi}|$)}  \\
    \hline
    \multicolumn{1}{c|}{\checkmark} & \multicolumn{1}{c|}{\checkmark} & \multicolumn{1}{c|}{\checkmark} & \multicolumn{1}{c|}{\checkmark} & 68.4  & \multicolumn{1}{c}{0.624} & \multicolumn{1}{c}{0.334}\\
   \multicolumn{1}{c|}{}  & \multicolumn{1}{c|}{\checkmark} & \multicolumn{1}{c|}{\checkmark}& \multicolumn{1}{c|}{\checkmark} & 67.8  & \multicolumn{1}{c}{0.706} & \multicolumn{1}{c}{0.354}\\
   \multicolumn{1}{c|}{}  &  \multicolumn{1}{c|}{}     & \multicolumn{1}{c|}{\checkmark} & \multicolumn{1}{c|}{\checkmark}& 67.1  & \multicolumn{1}{c}{0.751} & \multicolumn{1}{c}{0.368}\\
   \multicolumn{1}{c|}{} &   \multicolumn{1}{c|}{}    & \multicolumn{1}{c|}{}      & \multicolumn{1}{c|}{\checkmark} & 66.7  & \multicolumn{1}{c}{0.813} & \multicolumn{1}{c}{0.382} \\
   \multicolumn{1}{c|}{}  &  \multicolumn{1}{c|}{}     &  \multicolumn{1}{c|}{}     &   \multicolumn{1}{c|}{}    & 66.2    & \multicolumn{1}{c}{0.867} &  \multicolumn{1}{c}{0.395}\\
   \hline
    %\multicolumn{1}{l|}{} & \multicolumn{3}{c|}{VIT}     & 66.3    &  \multicolumn{1}{c}{1.030}\\
    %\bottomrule
   %\multicolumn{1}{l|}{} & \multicolumn{3}{c|}{Conv}  & 66.2    & \multicolumn{1}{c}{0.907} \\
   % \bottomrule
    \end{tabular}}%

  \label{tab:scale number}%
\end{minipage}
\end{minipage}

%---------------------------------------------------------------------------------------------------
\section{Conclusion}
\label{sec:conclusion}
In this paper, we proposed a novel Deformer module along with a multi-scale framework for unsupervised deformable registration. %To facilitate the mapping from image representation to spatial transformation and increase the interpretability of the network, 
The Deformer module was designed to formulate the prediction of displacement vector as the learning of most likely deformation directions and the offset length via attention mechanism. With the two fully connected layers applied on each pair of voxel-wise feature vectors independently, the Deformer module could better characterize the local spatial correlation with fewer parameters comparing with CNN or Transformer. Further, we showed that introducing the Deformer module along with auxiliary loss in a multi-scale manner to learn the displacement fields from coarse to fine could substantially boost the registration performance. Experiments on two publicly available datasets demonstrated that our strategy outperformed the traditional and learning-based benchmark methods. %A potential limitation of our method is that a rigid registration is needed for images with large misalignment and we only conduct experiments on the most common scenery of medical image registration.%, similar to most deep-learning-based approaches. 
%Besides, , i.e., human brain scans, evaluation on images with more flexible structures, such as abdomen or lung, will be performed in our future study.

\subsubsection{Acknowledgements.} 
This work was funded by the Scientific and Technical Innovation 2030-"New Generation Artificial Intelligence" (No.2020AAA0104100), Key R\&D Program of China (2018AAA0100104, 2018AAA0100100) and Natural Science Foundation of Jiangsu Province (BK20211164).
%Please place your acknowledgments at the end of the paper, preceded by an unnumbered run-in heading (i.e.3rd-level heading).

\bibliographystyle{splncs04}
\bibliography{egbib}

\clearpage
\title{Supplementary Materials}
\author{J. Chen, D. Lu, Y. Zhang, D. Wei, M. Ning, X. Shi, Z. Xu, and Y. Zheng}
\institute{The same with manuscript}
\maketitle
\begin{figure}[h] 
\centering
\includegraphics[width=\columnwidth]{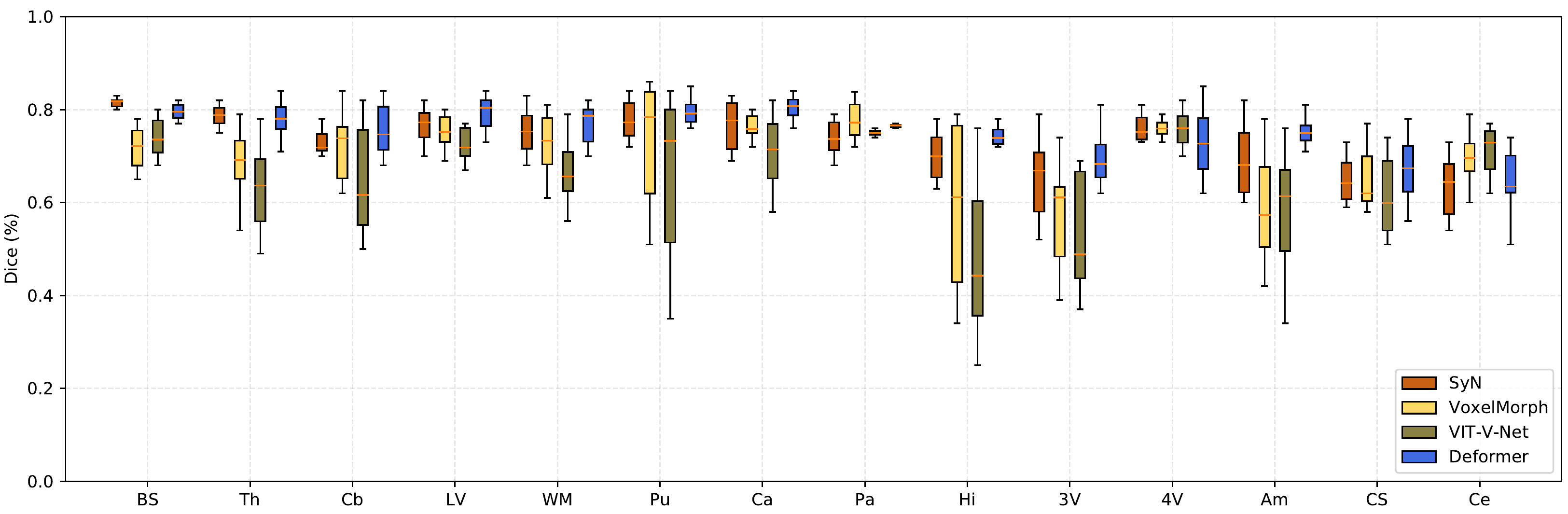} 
\caption{Boxplots of the average Dice scores of various anatomical structures, including the brain stem (BS), thalamus (Th), cerebellum cortex (Cb), lateral ventricle (LV), cerebellum white matter (WM), putamen (Pu), caudate (Ca), pallidum (Pa), hippocampus (Hi), 3rd ventricle (3V), 4th ventricle (4V), amygdala (Am), CSF (CS), and cerebral cortex (Ce), on Neurite-OASIS dataset for SyN, VoxelMorph, VIT-V-Net and our DMR method. The Dice for left and right hemispheres are averaged into one scores.}

\label{fig:boxplot} 
\end{figure}

\begin{figure}[h] 
\centering
\includegraphics[width=1\columnwidth]{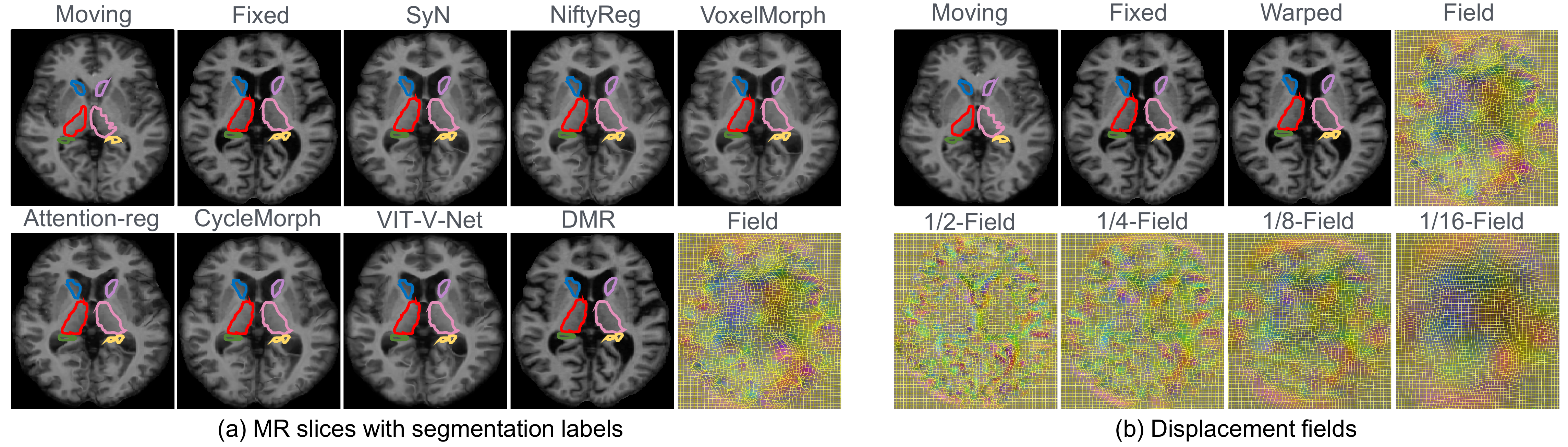} 
\caption{Qualitative results on Neurite-OASIS. (a) Segmentation labels of the example axial MR slices from the moving, fixed and warped images from different methods. The color curves represent the boundaries of several structures, including caudate (blue/purple), thalamus (red/pink), and hippocampus (green/yellow). (b) Visualization of displacement fields at different scales of our DMR method. The refining network can combine their information to enable high-resolution large-deformable registration.}
\label{fig:visual2} 
\end{figure}

\iffalse
\begin{figure}[t] 
\centering 
\includegraphics[width=\columnwidth]{Refine.pdf} 
\caption{The architecture of refining network. The fusion block is applied to combine the information from the displacement fields of neighbouring scales along with the feature maps of volume pairs.} %{\color{red} confusion, need to plot 'add' first then skip connection} 
\label{Fig.refine} 
\end{figure}
\fi

\begin{table}[htbp]
	\centering
	\caption{The impact of $K$ evaluated on LPBA40. When there is no multi-head mechanism ($K=1$), Dice score is obviously lower than others, while too much attention heads may cause overfitting and result in inferior performance.}
	\begin{tabular}{cccc}
	\toprule
	$K$     & Dice (\%) & \multicolumn{1}{c}{$|J_{\phi}|\leq0$ (\%)}     & \multicolumn{1}{c}{std($|J_{\phi}|$)} \\
	\midrule
	1     & 67.6  &    0.526   & 0.317  \\
	4    &  68.0  &   0.578    & 0.331  \\ 
	8    & 68.4  &    0.624   & 0.334  \\ 
	12   & 68.2     &  0.511     & 0.315 \\
	\bottomrule
	\end{tabular}
	\label{tab:head}%
\end{table}%

\begin{table}[htbp]
	\centering
	\caption{The impact of $N$ at each head evaluated on LPBA40. We can observe that with a large number of displacement basis, better performance can be achieved. Due to the limitation of GPU memories, we set the number $N$ as 64, which is sufficient to achieve the state-of-the-art performance.}
	\begin{tabular}{cccc}
	\toprule
	$N$     & Dice (\%) & \multicolumn{1}{c}{$|J_{\phi}|\leq0$ (\%)}     & \multicolumn{1}{c}{std($|J_{\phi}|$)} \\
	\midrule
	8     & 67.1  &    0.708   & 0.352  \\
	16    &  67.3  &   0.614    & 0.338  \\ 
	32    & 67.7  &    0.566   & 0.331  \\ 
	64   & 68.4     &  0.624     & 0.334 \\
	\bottomrule
	\end{tabular}
	\label{tab:number}%
\end{table}%

\iffalse
\begin{figure}[h] 
\centering
\includegraphics[width=\columnwidth]{conv block.pdf} 
\caption{The structure of the convolution block at each scale and the convolution head. (a) Each convolution block contains three convolutional layers, each of which has a stride of 1 followed by a group normalization and a ReLU activation function. (b) Each convolutional layer of the convolution head has a stride of 1, and an upsample layer is introduced after the second convolutional layer to restore the original resolution.}

\label{fig:conv block} 
\end{figure}
\fi

\begin{figure}[t] 
\centering 
\includegraphics[width=\columnwidth]{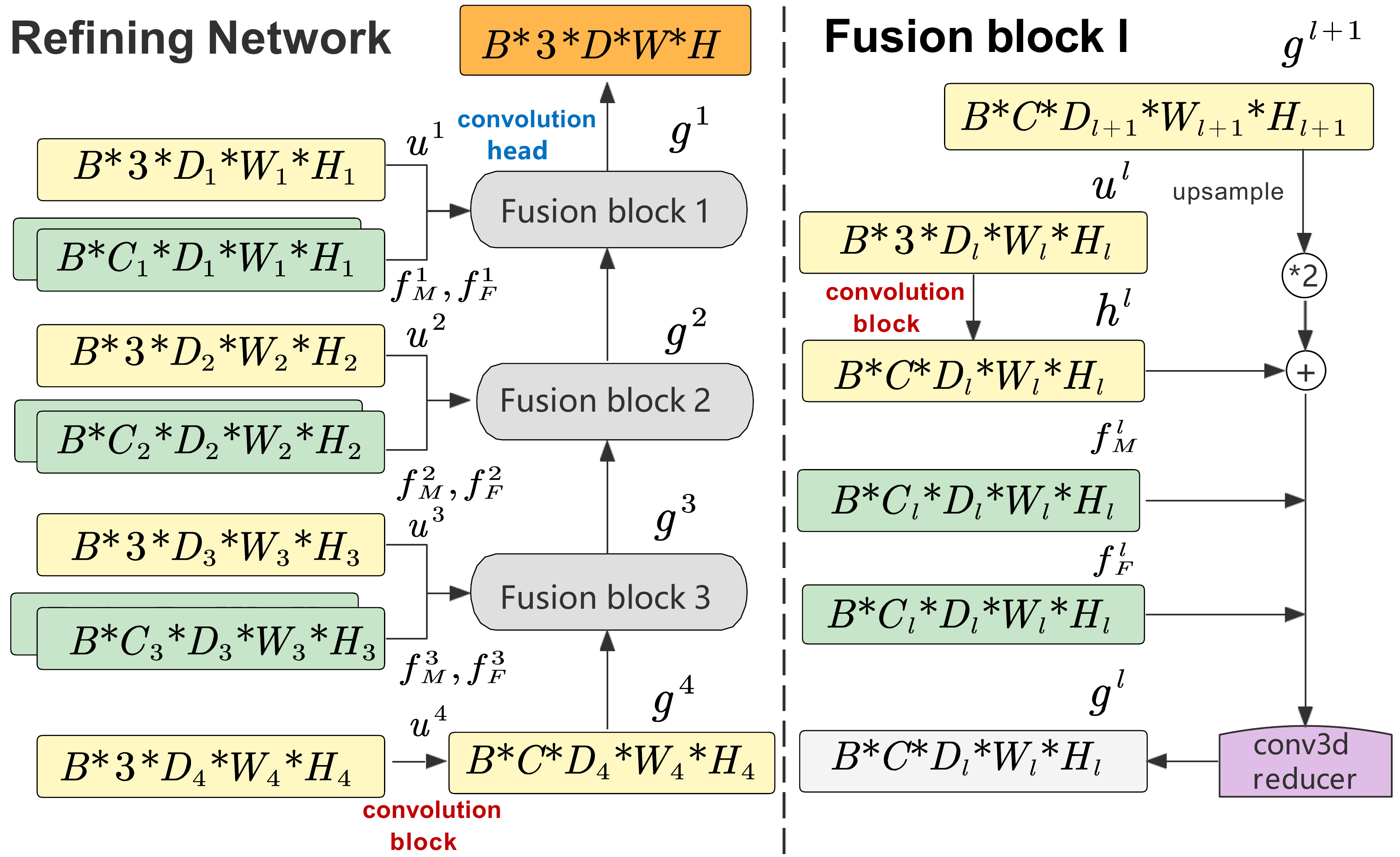} 
\caption{The architecture of refining network. The fusion block is applied to combine the information from the displacement fields of neighbouring scales along with the feature maps of volume pairs. Each convolution block contains three convolutional layers with 16, 64 and 128 channels (stride=1), respectively, each of which followed by a group normalization and a ReLU activation function. Note that the convolution blocks at different scales do not share weights and their kernel sizes are set as (5,5,5), (5,5,3), (5,3,3), (3,3,3) from large to small scale.} 
\label{Fig.refine} 
\end{figure}

\end{document}